\documentclass[12pt]{article}
\usepackage{amsfonts}
\usepackage{latexsym,cite,amssymb}
\usepackage{times}
\usepackage{graphicx}
\usepackage{color}
\def\be{\begin{eqnarray}}\def\ba{\begin{eqnarray}}
\def\ee{\end{endqnarray}}\def\ea{\end{eqnarray}}

\makeatletter
\renewcommand{\theequation}{\thesection.\arabic{equation}}
\@addtoreset{equation}{section} \makeatother
\input amssym.def
\input amssym.tex

\begin{document}

\title{\vskip -60pt
\vskip 20pt Acoustic black holes for relativistic  fluids~}
\author{ Xian-Hui Ge${}^{1}~$,
and  Sang-Jin Sin${}^{2}~$}
\date{}
\maketitle \vspace{-1.0cm}
\begin{center}
~~~
${}^{1}$Department of Physics, Shanghai University, Shanghai 200444, China\\
~{}
${}^{2}$Department of Physics, Hanyang University, Seoul 133-791, Korea\\
~{}
\\
{\small {E-mail: {{gexh@shu.edu.cn}, ~{sjsin@hanyang.ac.kr}}}}
~~~\\
~~~\\
\end{center}

\begin{abstract}
We derive a new acoustic black hole metric from the Abelian Higgs
model. In the non-relativistic limit, while the Abelian Higgs model
becomes the Ginzburg-Landau model, the metric reduces to an ordinary
Unruh type. We investigate the possibility of using (type I and II)
superconductors as the acoustic black holes. We propose to realize
experimental acoustic black holes by using spiral vortices solutions
from the Navier-stokes equation in the non-relativistic classical
fluids.
 \end{abstract}

{\small
\begin{flushleft}
\end{flushleft}}
\newpage

\tableofcontents 

\section{Introduction }
Black hole is characterized by the event horizon, a spherical
boundary out of which  even the light cannot escape. In 1974,
Hawking announced that black holes  are not black at all,
and emit thermal radiation at a temperature proportional to the horizon surface gravity. Although
several decades passed, the experimental verification of Hawking
radiation is still elusive.

Compared with the difficulties on the astrophysical side, analog
models of general relativity is more appealing since these models
are shedding light on possible experimental verifications of the
evaporation of black holes. In the remarkable paper of
Unruh\cite{unruh}, the idea of using hydrodynamical flows as analog
systems to mimic a few properties of black hole physics was
proposed. In this model, sound waves rather than light waves, cannot
escape from the horizon and therefore it is named ``acoustic (sonic)
black hole''. A moving fluid with speed exceeding the local sound
velocity through a spherical surface could in principle form an
acoustic black hole. The event horizon is located on the boundary
between subsonic and supersonic flow regions.

In general, any fluid flows could be the candidate of acoustic black
holes. Many classical and quantum systems have been investigated as
black hole analogues, including gravity wave \cite{unruhs},
water\cite{rou}, slow light\cite{leo1, leo2, leo3}, an optical
fiber\cite{leo4}, and an electromagnetic waveguide \cite{sch}.
 But considered the
detection of ``Hawking radiation" of acoustic black holes, quantum
fluid systems, rather than classical fluid systems are specifically
preferred in that Hawking radiation is actually a quantum
phenomenon. Up to now, superfluid helium II \cite{volv}, atomic
Bose-Einstein condensates \cite{garay}, one-dimensional
Fermi-degenerate noninteracting gas \cite{gio} have been proposed to
created an acoustic black hole geometry in the laboratory. The
possible application of acoustic black holes  to quantum information
theory has been suggested in \cite{ge}.

The first experimental realization of acoustic black hole reported
was conducted in a Bose-Einstein condensate \cite{laha}. So far it
has been assumed that the  system is non-relativistic, because all
the atomic system is likely in such regime. However, some strongly
interacting system shows some evidences for exotic behavior  that
the matter is based on the heavy atoms but the basic excitations has
massless character \cite{Sin,Gubser}. In fact the linear and
massless dispersion relation is canonical for fermionic system.
 Therefore  it is
interesting to ask what happen to the acoustic black hole
 for the relativistic fluid system.

In this paper, we explore a relativistic version of acoustic black
holes from the Abelian Higgs model. Another motivation comes from
the feasibility of detecting Hawking temperature associated with the
acoustic black hole. Since the Hawking temperature depends on the
gradients of flow speed at the horizon,  detecting thermal phonons
radiating from the horizons is very difficult. In fact the Hawking
temperature calculated from models in Bose-Einstein condensates so
far is very low ($\sim$ nano Kelvin).  {In constrast,  the
acoustic black holes created in the color superconducting phase in
the dense quark matter and the pion superfluid phase at finite
isospin density
  may provide us  examples of relativistic superfluid and higher temperature acoustic black hole}.  The line elements of the black hole metric we obtained are found
to be different from the non-relativistic version, which has been
widely used for almost 30 years.

\section{Acoustic black holes from Abelian Higgs model}
The Abelian Higgs model is the Mexican-hat model coupled to
electromagnetism with the action
\begin{equation}\label{higgs}
S=\int d^4x
(-\frac{1}{4}F^{\mu\nu}F_{\mu\nu}+|(\partial_{\mu}-ieA_{\mu})\phi|^2+m^2
|\phi|^2-b |\phi|^4),
\end{equation}
where $F_{\mu\nu}=\partial_\nu A_\mu-\partial_\mu A_\nu$. The Planck
units $\hbar=c=k_B=1$ are used here. We will turn on the Planck
constant $\hbar$ in section 3. The corresponding equations of motion
can be deduced from the action with one governing on the
electromagnetic field and the other on the scalar field.  We write
down only the second equation of motion for the purpose of deriving
the acoustic black hole geometry
\begin{equation}\label{higgs}
\square\phi+2 i e A_{\mu}\nabla^{\mu}\phi-e^2 A_{\mu}A^{\mu}\phi+m^2
\phi-b \mid\phi\mid^{2}\phi=0,
\end{equation}
in the $\partial^{\mu} A_{\mu}=0$ gauge.

 Multiplying (\ref{higgs}) with $\phi^{*}$, we obtain
\begin{equation}
\label{41}
\phi^{*}\left(-\partial_{t}\partial^{t}\phi+\nabla_{i}\nabla^{i}\phi+2
i e A_{\mu}\nabla^{\mu}\phi-e^2 A_{\mu}A^{\mu}\phi\right)+m^2\phi
\phi^{*}-b \mid\phi\mid^{4}=0,
\end{equation}
where the index $\mu=t, x_1...x_3$ and $i=x_1...x_3$. We can obtain
another equation by multiplying the conjugation of (\ref{higgs})
with $\phi$
\begin{equation}
\label{42}
\phi\left(-\partial^{2}_{t}\phi^{*}+\nabla_{i}\nabla^{i}\phi^{*}+2 i
e A_{\mu}\nabla^{\mu}\phi^{*}-e^2
A_{\mu}A^{\mu}\phi^{*}\right)+m^2\phi \phi^{*}-b \mid\phi\mid^{4}=0.
\end{equation}
The combination of (\ref{41}) and (\ref{42}) yields two equations
\begin{equation}
-\partial_{t}\left(\phi^{*} \partial^{t}\phi-\phi
\partial^{t}\phi^{*}\right)+\nabla_{i}\left(\phi^{*} \nabla^{i}\phi-\phi
\nabla^{i}\phi^{*}\right)+2 ieA_{\mu}\left(\phi^{*}
\nabla^{\mu}\phi+\phi \nabla^{\mu}\phi^{*}\right)=0.
\end{equation}
and
\begin{eqnarray}
&&-\partial_{t}\left(\phi^{*} \partial^{t}\phi+\phi
\partial^{t}\phi^{*}\right)+ 2 |\partial_{t}\phi|^{2}+\nabla_{i}\left(\phi^{*} \nabla^{i}\phi+\phi
\nabla^{i}\phi^{*}\right)-2 |\partial_{i}\phi|^2+2 m^2 |\phi|^2
\nonumber\\  &&+2 ieA_{\mu}\left(\phi^{*} \nabla^{\mu}\phi-\phi
\nabla^{\mu}\phi^{*}\right)-2b |\phi|^4-2e^2
A_{\mu}A^{\mu}|\phi|^2=0.
\end{eqnarray}
With the assumption $\phi=\sqrt{\rho(\vec{x},t)}e^{i
\theta(\vec{x},t)} $, the above equations reduce to
\begin{equation}
\label{main1} -\partial_{t}\left[\rho
(\dot{{\theta}}-eA_{t})\right]+\nabla_{i}\left[\rho\left(
\nabla^{i}\theta +eA_{i}\right)\right]=0,
\end{equation}
\begin{equation}
\label{main2} \frac{\square
\sqrt{\rho}}{\sqrt{\rho}}+(\dot{\theta}-eA_{t})^2-(\nabla_{i}\theta+eA_{i})^2
+m^2 - b \rho=0,
\end{equation}
where (\ref{main1}) is the continuity equation and (\ref{main2})
looks like an equation describing hydrodynamical fluid with an extra
quantum correction term $\frac{\square \sqrt{\rho}}{\sqrt{\rho}}$,
which is the so called quantum potential. (\ref{main1}) and
(\ref{main2}) are actually relativistic and thus are Lorentz
invariant. This is a crucial difference from previous literature
where only the Galilean invariant fluid equations  were  considered.
Hydrodynamics is  a long wavelength effective theory for
physical fluids. Quantum gravity, or even very exotic field
theories, when heated up to finite temperature, also behave
hydrodynamically at long-wavelength  and long-time scales.
 Now we consider perturbations around the
background $(\rho_0, \theta_0)$: $\rho=\rho_0+\rho_{1}$ and
$\theta=\theta_0+\theta_1$, and rewrite (\ref{main1}) and
(\ref{main2}) as
\begin{eqnarray}
\label{continue} &&-\partial_{t}\left[\rho_0
\dot{\theta}_{1}+(\dot{\theta}_{0}-eA_{t})\rho_1\right]+\nabla\cdot\left[\rho_0
{\nabla\theta}_{1}+\rho_1 (\nabla{\theta}_{0}+eA_{i})\right]=0, \\
\label{Euler}&&2(\dot{\theta}_{0}-eA_{t})\dot{\theta}_1-2(\nabla{\theta}_{0}+eA_{i})\cdot
\nabla\theta_{1}-b \rho_{1}+D_{2}\rho_{1}=0,
\end{eqnarray}
where we have defined
$$D_{2}\rho_{1}=-\frac{1}{2}\rho^{-\frac{3}{2}}_{0}(\square\sqrt{\rho_0})\rho_{1}
+\frac{1}{2}\rho^{-\frac{1}{2}}_{0}\square(\rho_0^{-\frac{1}{2}}\rho_1).$$
In this work, we do not consider   fluctuations of the electromagnetic field $A_{\mu}$. It is a background field that can affect  the acoustic black hole
geometry.  The equations (\ref{continue}) and (\ref{Euler})
corresponds to a  relativistic  fluids equation if
    $b\rho_1$ term and $D_2\rho_1$ (the quantum potential) term  are absent.
     Actually, the
$D_2$ term  contains the second derivative of slowly varying $\rho$,
which can be negligible in the  hydrodynamic region where $k,\omega$ are small.
  We will turn on the quantum potential term only in next subsection.
 In fact it is  well-known that  in the Madelung representation of the condensation,
$$\phi=\sqrt{\rho(\vec{x},t)}e^{i \theta(\vec{x},t)},$$  the
Schrodinger equation can be rewritten as a continuity equation plus
an Euler equation for an irrotational and inviscid fluid when the
quantum potential can be neglected. The Abelian Higgs model here
shows that the same property holds for relativistic fluids.

 For the simplicity of
calculation, let us introduce variables $\omega_0$ and $\vec{v}_0$
with the definition $$\omega_0=-\dot{\theta}_{0}+eA_{t} {\rm ~~and~~}
 \vec{v}_0=\nabla{\theta}_{0}+e\vec{{A}}.$$ We then obtain
\begin{eqnarray}
\label{main3} &&-\partial_{t}\left[\rho_{0}\dot{\theta}_{1}
+\frac{2}{b}\left(\omega^2_0\dot{\theta}_1+\omega_0\vec{v}_0
\cdot \nabla\theta_1\right)\right]\nonumber\\
&&+\nabla\cdot\left[\rho_{0}\nabla\theta_{1}+\frac{2}{b}\left(-\vec{v}_0
\omega_{0}\dot{\theta}_{1}- \vec{v}_0\cdot \nabla\theta_1
\vec{v}_0\right)\right]=0.
\end{eqnarray}
Comparing the above equation with the massless Klein-Gordon equation
\begin{equation}
\frac{1}{\sqrt{-g}}\partial_{\nu}(\sqrt{-g}g^{\mu\nu}\partial_{\mu}\theta_1)=0,
\end{equation}
one may find
\begin{equation}
\sqrt{-g}g^{\mu\nu}\equiv  \left[ \matrix{-\frac{b}{2}
\rho_0-\omega^2_0&\vdots&-v_0^j \omega_0\cr
               \cdots\cdots&\cdot&\cdots\cdots\cdots\cdots\cr
           -v_0^i \omega_0&\vdots&( \frac{b}{2} \rho_0\delta^{ij} - v_0^i v_0^j )\cr }
\right]. \label{E}
\end{equation}
Defining the local speed of sound
\begin{equation}
\label{cs} c^2_{s}=\frac{b}{2 \omega^2_0}\rho_0,
\end{equation}
  the metric for acoustic black hole  yields the form
\begin{equation}
g_{\mu\nu}\equiv
\frac{\frac{b\rho_0}{2c_s}}{\sqrt{1+c^2_s-v^2}} \left[ \matrix{-(c^2_s-v^2)&\vdots&-v^j \cr
               \cdots\cdots&\cdot&\cdots\cdots\cdots\cdots\cr
           -v^i &\vdots&( 1+c^2_s-v^2)\delta^{ij} + v^i v^j \cr }
\right], \label{remetric}
\end{equation}
where we made replacement $v_0^i\rightarrow v^i \omega_0$ and  notation
 $v^2=\sum_{i}v^i v^i$. We now have a relativistic version of
acoustic black hole which is a generalization of \cite{unruh}.
It is worth noting that the sound velocity $c_s$ is a function of the electromagnetic field $A_t$, which is  treated as a slowly varying quantity.

In the non-relativistic limit, $c_s\ll
1,\vec{v}\ll 1$, and keeping only the leading order term of each
line element, we can recover the metric from non-relativistic theory
first obtained by Unruh \cite{unruh}, but with an overall factor
difference
\begin{equation}
\label{un}
 g_{\mu\nu}\equiv
\left(\frac{\rho_0}{c_s}\right) \left[
\matrix{-(c^2_s-v^2)&\vdots&-v^j \cr
               \cdots\cdots&\cdot&\cdots\cdots\cdots\cdots\cr
           -v^i &\vdots&\delta^{ij}  \cr }
\right]. \label{nonrelativistic}
\end{equation}
Equivalently, we can rewrite (\ref{remetric}) as
\begin{eqnarray}
\label{m2}
ds^2&=&\frac{b}{2}\frac{\rho_0}{c_s}\frac{1}{\sqrt{1+c^2_s-v^2}}\left[-(c^2_s-v^2)dt^2-2
\vec{v}\cdot d\vec{x}dt+(\vec{v}\cdot
d\vec{x})^2+(1+c^2_s-v^2)d\vec{x}^2\right]\nonumber\\
&=&\frac{b}{2}\frac{\rho_0}{c_s}\frac{1}{\sqrt{1+c^2_s-v^2}}\left[-(c^2_s-v^2)d\tau^2
+(1+c^2_s-v^2)\left(\frac{{v}^i {v}^j
}{c^2_s-v^2}+\delta_{ij}\right)dx^i dx^j\right],
\end{eqnarray}
where we have defined a new time coordinate by
$d\tau=dt+\frac{\vec{v}\cdot d\vec{x}}{c^2_s-v^2}$. In this
coordinate the acoustic geometry is actually stationary.

In the case $v_z=0$ , but $v_r\neq 0$ and $v_{\phi}\neq 0$, we can
simplify the metric in a Kerr-like form
\begin{eqnarray}
\label{burgers}
ds^2&=&{\frac{b}{2}\frac{\rho_0}{c_s}}\left[-N^2(v_r,v_{\phi})d\tau^2
+\frac{1}{N^2(v_r,v_{\phi})}dr^2 +\sqrt{1+c^2_s-v^2}dz^2
\right.\nonumber\\&&\left. +N^2(v_r,v_{\phi})r^2
d\varphi^2+\frac{(v_{\phi} d\tau-rd\varphi)^2}{\sqrt{1+c^2_s-v^2}}
\right],
\end{eqnarray}
where
\begin{equation}
N^2(v_r,v_{\phi})=\frac{c^2_s-v^2_r}{\sqrt{1+c^2_s-v^2}},
\end{equation}
and the coordinate transformations have been used
\begin{eqnarray}
dt = d\tau +\frac{v_{r}}{{v_r}^2-c^2_s} dr,~~~~~~
 d\phi = d\varphi + \frac{v_r v_{\phi}}{{v_r}^2-c^2_s}\frac{dr}{r}.
\end{eqnarray}
 When $v_z\neq0$, the metric becomes non-spherically symmetric
and non-axisymmetric. In non-relativistic limit, the geometry
(\ref{burgers}) reduces to a ``draining bathtub'' metric first
appeared in \cite{visser}.

At the region far from the acoustic horizon, the normal modes of the
$\theta_1$ field are purely outgoing. The operator $\theta_1$ can be
expanded in terms of modes in the region outside the horizon. An
observer traveling with the fluid as it flows through the acoustic
horizon will see  that the field $\theta_1$ is essentially the
vacuum state in analogy with the behavior of the normal modes of a
scalar field in Schwarzschild coordinates by a freely falling
observer \cite{unruh}.
 The corresponding Hawking temperature at the event horizon reads
\begin{equation}
T_{H}=\frac{\hbar}{2 \pi k_B }\frac{\partial (c_s-v^{i})}{\partial
x_i}|_{\rm horizon}.
\end{equation}
The relativistic version of Hawking temperature from Abelian Higgs
model is identical with the non-relativistic one.

 Since there is no real gravity, the Newton constant
$G_{N}$ is not involved in the formulation of the acoustic ``surface
gravity'' and Hawking temperature.  One may  compare the acoustic
Hawking temperature with Unruh temperature in Rindler coordinates.
The acoustic temperature and the Unruh temperature yield almost the
same form: The acoustic temperature is proportional to the gradients
of the fluid kinetic energy (i.e. $\propto\partial_i
(c^2_s-v^2_i)|_{\rm horizon}$ ) at the horizon, while the Unruh
temperature ($T=\frac{{a_R}}{2 \pi}$), observed by an accelerating
observer is proportional to the acceleration $a_R$ ($=\frac{\partial
v}{\partial t}$). According to Einstein equivalence principle, no
experiment can distinguish the acceleration due to gravity from the
inertial acceleration due to a change of velocity. Thus, the Unruh
effects in Rindler coordinates and Hawking effects in astrophysical
black hole spacetime are equivalent. It is surprising that the
acoustic geometry provides us another way to realize the
equivalence principle by the relation {(Acceleration) = (Intensity
of the gravitational field)=(Gradients of the kinetic energy)}. One
point that should be made clear is that  acoustic analog  models of
black holes could be used to mimic the kinematical but not the
dynamical aspects of gravitational systems. For example, it is
impossible to reproduce black hole thermodynamics using acoustic
analogue models of black holes\cite{visser1}.

\subsection{Dispersion relation}
 We have shown that by studying the
linearized  equations of motion of the complex scalar fields of Abelian Higgs model,
the perturbations obey a Klein-Gordon  equation in the curved space-time.
Here we stress that
since we worked in the  long-wavelength and  slowly varying limit,
the quantum potential term (the $D_2$ term) is  neglible.

In  the high-momentum or eikonal regime,  the contribution of the quantum potential would be important.
In \cite{b}, the authors explored the dispersion relation with the
contribution of the quantum potential. In our case, both the quantum
potential and the relativistic effects will contribute to the
dispersion relation.   IN this regime,  the phase fluctuation $\theta_1$
and the density fluctuation $\rho_1$ can be treated as a function
with a slowly varying amplitude and  rapidly varying phase:
\begin{eqnarray}
&&\theta_1 \sim \rho_1\sim \rm Re\{{\rm exp(-i \Omega t+i
\vec{k}\cdot
\vec{x})}\}\\
&&\Omega=\frac{\partial_t \theta_1}{\theta_1}=\frac{\partial_t
\rho_1}{\rho_1},~~~ k_i=-\frac{\nabla_i \theta_1}{\theta_1}=-\frac{\nabla_i \rho_1}{\rho_1}
\end{eqnarray}
The operator $D_2$ can be approximately written as
\begin{eqnarray}
D_2 \rho_1
&\equiv&-\frac{1}{2}\rho^{-\frac{3}{2}}_{0}(\square\sqrt{\rho_0})\rho_{1}
+\frac{1}{2}\rho^{-\frac{1}{2}}_{0}\square(\rho_0^{-\frac{1}{2}}\rho_1)\nonumber\\
&= &\frac{\rho_1 (\partial_\mu
\rho_0)^2}{2\rho^3_0}-\frac{\partial_{\mu}\rho_0\partial_{\mu}\rho_1}{2\rho^2_0}
-\frac{\rho_1\square \rho_0}{2\rho_0^2}+\frac{\square \rho_1}{2 \rho_0}\nonumber\\
&\approx & \frac{\square \rho_1}{2 \rho_0}\nonumber\\&=&
-\frac{1}{2}\rho^{-1}_0 (k^2-\Omega^2)\rho_1,
\end{eqnarray}
 where $\mu=t,x_1...x_3$.
We have similar result for $D_2 \theta_1$. Therefore,  we can
treat  $D_2=-\frac{1}{2}\rho^{-1}_0 (k^2-\Omega^2)$ in the eikonal regime.
The Klein-Gordon equation (\ref{continue}) and (\ref{Euler}) become
\begin{equation}
\label{dispersion} \frac{D_2-b}{2}\rho_0(k_i k_j
\delta^{ij}-\Omega^2)-(\omega_0 \Omega+v^i_0k_j)^2=0.
\end{equation}
With the expression of $D_2$ above,  Eq.(\ref{dispersion}) becomes
\begin{equation}
\label{dispersion2} \frac{b\rho_0}{2}\left(k^2-\Omega^2\right)-
\left(\omega_0\Omega+v^i_0k_i\right)^2+\frac{1}{4}\left(k^2-\Omega^2\right)^2=0.
\end{equation}
 {In the eikonal limit, the last term in the above equation
will become dominant and we can neglect the other two terms, which
gives }
\begin{equation}
\Omega=k.
\end{equation}

 On the other hand,   in the hydrodynamics regime, we can neglect
the $D_2$ term again  in the left hand of
Eq.(\ref{dispersion}). To be definite, we choose  $k^i=k\delta^{i1}$, and define
$v^i =v^i_0 /\omega_0$. Then we easily obtain the dispersion relation
\begin{equation}
\label{dispersion3}\Omega=\frac{-v^1+
c_s\sqrt{1+c^2_s-(v^1)^2}}{1+c^2_s}k.
\end{equation}
The dispersion relation  is quiet different from the result  of
\cite{visser}. For large wavelength and in the non-relativistic
limit, it reads
\begin{equation}
\label{dispersion4}\Omega = (c_s-v^1)(1+(c_s+v^1)/2)k.
\end{equation}
 The sound velocity  in hydrodynamic limit has a correction
to the familiar  value $c_s$. The main correction is simply due to the fluid motion
and the next term is higher order. Since  $v^1$, the generalized physical velocity in the presence of the electro-magnetic fields,
depends on background electric and magnetic fields, as one can see from
\begin{equation}
\label{dispersion5}v^1=(\partial_x \theta_0 +eA_x)/(-{\dot \theta}+eA_t) ,
\end{equation}
so is the sound velocity.

We stress that we assume that the back ground fields
change very little within a wave length.
If $c^2_s$ and $v^2$ can be neglected, compared with the speed
of light,   then the relation return to a linear dispersion
relation $\Omega \approx c_s |k|$.

\section{Non-relativistic limit: Acoustic black holes in superconductors}
  In the non-relativistic
limit, the Abelian Higgs model reduces to the Ginzburg-Landau theory
\begin{equation}
i\hbar \frac{\partial}{\partial t}\psi(\vec{r},t)= -\frac{\hbar
^2}{2
m^{*}}\vec{\mathfrak{D}}^2\psi(\vec{r},t)+a(T)\psi(\vec{r},t)+b(T)\mid\psi(\vec{r},t)\mid^2
\psi(\vec{r},t), \label{GL}
\end{equation}
where $\vec{\mathfrak{D}}=\nabla+\frac{2ie}{\hbar}\vec{A}$,  $m^*$
is the mass of a cooper pair, $a(T)$ and $b(T)$ are two parameters
that depend on temperature. We have turned on the Planck constant
$\hbar$. The Ginzburg-Landau coherence length is defined by
\begin{equation}
\xi(T)=\left(\frac{\hbar^2}{2m^{*}\mid a(T)\mid}\right)^{1/2}.
\end{equation}
 Assuming
$\psi(\vec{r},t)=\sqrt{\rho(\vec{r},t)}e^{i\theta(\vec{r},t)}$,  we
have
\begin{eqnarray}
&&\frac{\partial \rho }{\partial t}+\vec{\nabla}\cdot (\rho
\vec{v})=0,
\label{rho} \\
&&\hbar \frac{\partial \theta}{\partial t}=\frac{\hbar^2}{2
m^{*}}\frac{\nabla^2
\sqrt{\rho}}{\sqrt{\rho}}-\frac{m^{*}}{2}\vec{v}^2-a(T)-b(T)\rho.\label{rho2}
\end{eqnarray}
where we have used the gauge $\nabla\cdot\vec{A}=0$, and the
velocity is denoted by $\vec{v}=\frac{\hbar
\nabla\theta}{m^{*}}+\frac{2e}{m^{*}}\vec{A}$.  The first term in
the right hand of (\ref{rho2}) corresponds to the quantum potential.
The above two equations are completely equivalent to the
hydrodynamical equations for irrotational and inviscid fluid apart
from the quantum potential. In the long-wavelength approximation,
the contribution coming from the linearization of the quantum
potential can be neglected.

Linearizing Eqs.(\ref{rho}) and (\ref{rho2}) around the
background ($\rho_0, \theta_0$), we get
\begin{eqnarray}
\frac{\partial \rho_1}{\partial t}+\nabla\cdot
\left(\rho_1\vec{v}_0+\rho_0
\nabla\theta_1\right)=0,\\
\frac{\partial \theta_1 }{\partial t}+\vec{v}_0\cdot\nabla
\theta_1+\frac{b(T)}{m^{*}}\rho_1=0.
\end{eqnarray}
Eliminating $\rho_1$, the above equations lead  to
\begin{equation}
\label{KG} \frac{\partial}{\partial
t}\left(-\frac{m^{*}}{b(T)}\frac{\partial \theta_1 }{\partial
t}-\frac{m^{*}}{b(T)}\vec{v}_0\cdot\nabla\theta_1\right)+\nabla\cdot\left(\rho_0\nabla\theta_1
-\frac{m^{*}}{b(T)}\vec{v}_0\frac{\partial\theta_1}{\partial
t}-\frac{m^{*}\vec{v}_0}{b(T)}\vec{v}_0\cdot\nabla\theta_1\right)=0,
\end{equation}
where $\vec{v}_0=\frac{\hbar
\nabla\theta_0}{m^{*}}+\frac{2e}{m^{*}}\vec{A}$. One can extract a
metric from (\ref{KG}) because it has the same form as a  massless
Klein-Gordorn equation in curved space-time. The non-relativistic
version of acoustic metric from (\ref{KG}) is read as
\begin{equation}\label{metric}
ds^2=\frac{\rho_0}{m^* c_s}\bigg\{-(c^2_s-v^2_0)dt^2-2\vec{v}_0\cdot
d\vec{r}dt+d\vec{r}\cdot d\vec{r}\bigg\},
\end{equation}
where $c_s=\sqrt{\frac{b(T)}{m^{*}}\rho_0}$ is denoted as the
``sound velocity''. Actually, one can write the sound velocity in
terms of the coherent length
\begin{equation}\label{sound}
c_s=\frac{\hbar}{\sqrt{2}m^{*} \xi(T)}.
\end{equation}
One should not identify the ``sound velocity'' defined here with the
sound velocity produced by the crystal lattices. Here $c_s$ is a
function of the density $\rho_0$ of superconducting electrons  and
temperature-dependent parameter $b(T)$.

We would like to do some calculations on the necessary conditions
for the formation of an acoustic event horizon in Ginzburg-Landau
theory. We will show that it would be a tough task to realize
acoustic black holes by using type I and type II superconductors.
The definition of acoustic black hole requires $\frac{v_0}{c_s}> 1$
in some region. From the London equation
\begin{equation}
\label{london} \nabla^2 {\bf{\vec{ B}}}-\frac{1}{\lambda (T)}{\bf
\vec{B}}=0,
\end{equation}
and the Maxwell equation
\begin{equation}
\label{maxwell} \nabla \times \bf {\vec{ B}}=\mu_0 \bf {\vec{j}}_s,
\end{equation}where $\lambda (T)=(\frac{m_e}{\mu_0 n_s e^2})^{1/2}$ and $\bf \vec{j}_s$ denote the penetration
depth and the supercurrent, respectively. We
have a finite magnetic field , say ${\bf \vec{B}}=(0,B_y(x),0)$,
outside a superconducting region. After
simple computation, we find the supercurrent yields \cite{poole}
\begin{equation}
j_{sz}= \frac{H(T)}{\lambda(T)}.
\end{equation}
The critical current density above is only valid for clean type I
superconductors. On the other hand, the current density has the form
\begin{equation}
\label{v1} j_{sz}=e n_s v_{0s},
\end{equation}
where $n_s=2|\psi^2|=2\rho_0$ denotes the superfluid density, and
$v_{0s}$ is the velocity of the Cooper pairs. We then have
\begin{equation}
\label{ge}
 \frac{v_{0s}}{c_s}=\frac{\sqrt{2}\kappa H(T)}{H_{c2}}=\frac{H(T)}{H_{c}},
\end{equation}
where $\kappa=\frac{\lambda(T)}{\xi(T)}$ , and $H_{c2}=\frac{h}{4\pi
e\mu_0{\xi^2(T)}}=\sqrt{2}\kappa H_c$. In the standard textbook,
$\kappa$ denotes the G-L parameter, which describes the difference
between type I and type II superconductors:
$\kappa<\frac{1}{\sqrt{2}}$ is for  type I, and $\kappa
>\frac{1}{\sqrt{2}}$ is for type II.  $H_{c2}$ denotes the critical
magnetic field,  and $H_{c}$ the thermodynamic critical field. From
(\ref{ge}), we can see that for type I superconductors, when
$v_{0s}=c_s$, the superconducting phase is broken and return to
Normal states.
Now we are going to solve (\ref{london}) in the high $\kappa$ limit,
$\lambda\gg \xi$, which is valid for the cooper-oxide
superconductors. We will show that even near the normal core of the
vortex, it is hard for the speed of superconducting electrons to
exceed  the ``sound velocity''-- $c_s$. For simplicity, the vortex
is assumed to be infinitely long and axially symmetric so that there
are no $z$ or angular dependence of its field distribution. We then
write Eq.(\ref{london}) in cylindrical coordinates
\begin{equation}
\frac{1}{r}\partial_r(r\partial_r\bf{B})-\frac{1}{\lambda^2}
\bf{B}=0.
\end{equation} This solution is given by \cite{poole}
\begin{equation}
B(r)=\frac{\phi_0}{2\pi\lambda^2}K_0(r/\lambda),
\end{equation} where $K_0(r/\lambda)$ is the zeroth-order modified
Bessel function. The current density is derived by the Maxwell
equation (\ref{maxwell}),
\begin{equation}
\vec{j}_s(r)=\frac{\phi_0}{2\pi\mu_0\lambda^3}K_1(r/\lambda)
~\bf{\vec{e}}_{\phi}
\end{equation}
We are interested in the asymptotic behaviors at small radial
distances $r \ll \lambda$, where $K_1(r/\lambda) \approx
\frac{\lambda}{r}$. The speed of Cooper pairs is then given by
\begin{equation}
\label{v2} v_{0s}=\frac{j_s(r)}{n_s e}
\end{equation}
We then find
\begin{equation}\label{rc}
\frac{v_{0s}}{c_s}=\frac{\sqrt{2} \xi}{r}.
\end{equation}
The formation of acoustic black holes requires ${v_{0s}}/{c_s}>1$,
i.e. $r<\sqrt{2} \xi${\footnote{The order parameter also reaches
its maximum at $r=\sqrt{2} \xi$ \cite{poole}.}}.  {But at the core
region of the vortex $r<  \xi$, the order parameter and the
superconducting current would decay as $r\rightarrow 0$, since it
corresponds to a normal state. (\ref{v1}) tells us that for fixed
electron velocity ${v_{0s}}$,   the current $j_{sz}$ decreases as
$n_s(=2\rho_0)$ decreases. Since the ``sound velocity'' $c_s$
depends on the density $\rho_0$, we expect that in the region $\xi
<r<\sqrt{2} \xi$, electron velocity can exceed the sound velocity.
However, it would be interesting if one can verify this point
experimentally.}

In order to trap phonons, one need input a sink so that electrons
can propagate along $\phi$- and $r$-direction simultaneously. For
this purpose, we need turn to spiral flux vortices in the type II
superconductors. The spiral flux vortices \cite{clem} in
current-carrying type-II superconducting cylinders subjected to
longitudinal magnetic fields can be used to mimic acoustic black
holes. Consider  a type-II superconducting cylinder of radius $a$
containing a spiral vortex, in the presence of a longitudinal
applied magnetic field $H_{\alpha}$ and transport current $I$, where
$H_{\alpha}$ and $I$ are positive in the $z$ direction. The Lorentz
force imposed on the vortex were found to be radially inward. Thus
the spiral vortex has a tendency to move to the center of the
cylinder. One can determine the terminal speed $v^{r}_{0s}$ of the
contracting vortex by balancing the Lorentz force and the viscous
drag force $\eta v^{r}_{0s}$, where $\eta$ is a phenomeno logical
viscous drag coefficient per unit length of vortex \cite{clem}.

Actually, the spiral vortex in the presence of a sufficiently large
current density parallel to its axis, is unstable against a helical
perturbation \cite{clem}. But this does not prevent the formation of
acoustic black hole geometry. In the next section, we investigate
the spiral vortex geometry from classical hydrodynamics in the
non-relativistic frame and give the exact metric for a spiral-vortex
geometry.

\subsection{Spiral-Vortex geometry}
In \cite{visser}, the author constructed a ``draining bathtub''
black hole metric by a $(2+1)$ dimensional Newtonian flow with a
sink at the origin. An stable vortex solution to the Navier-Stokes
equation requires each component of the fluid velocity to be
non-vanishing. Such an exact solution to Navier-Stokes equation was
first found by Sullivan\cite{sullivan} and later observed in
experiments\cite{leslie}. Now, we utilize the Sullivan vortex
solution of Navier-Stokes equation to model a 4-dimensional acoustic
black hole metric, which could be realized in experiments. The
non-relativistic Navier-Stokes equation is given by
\begin{equation}
\partial_t \vec{v}+\vec{v}\cdot \nabla \vec{v}=-\vec{\nabla}p/\rho+\nu
\nabla^2 \vec{v}+\vec{f}/\rho,
\end{equation}
 where $\vec{v}$ is the fluid velocity, $p$ the fluid pressure, $\rho$ the mass density,  $\nu$ the
shear viscosity and $\vec{f}$ an externally specified forcing
function. The continuity equation generally accompanies the Navier-Stokes equation
\begin{equation}
\partial_t \rho+\nabla \cdot (\rho \vec{v})=0.
\end{equation}
By defining $\vec{v}=\nabla \psi$, the  Navier-Stokes equation can be rewritten as
\begin{equation}
\partial_t \psi+\frac{1}{2}(\nabla \psi)^2+\int^{p}_0\frac{dp'}{\rho}-\nu\nabla^2\psi=0,
\end{equation}
where we have chosen vanishing external force $\vec{f}=0$. Now linearize these
equations around a fixed background $(\rho_0,p_0,\psi_0)$ and set $\rho=\rho_0+\rho_1$, $p=p_0+p_1$, and
$\psi=\psi_0+\psi_1$. After linearizing the equation of motion and the continuity equation,
the equations for $\rho_1$ and $\psi_1$ yield
\begin{eqnarray}
\label{co}
\frac{\partial \rho_1}{\partial t}+\nabla\cdot
\left(\rho_1 \vec{v}_0+\rho_0
\nabla\psi_1\right)=0,\\
\label{ns}
\frac{\partial \psi_1 }{\partial t}+\vec{v}_0\cdot\nabla
\psi_1+\frac{p_1}{\rho_0}-\nu \nabla^2\psi_1=0.
\end{eqnarray}
One can further assume the fluid is barotropic, which implies that $\rho$ is a function of $p$ only
\footnote{This assumption was first used implicitly in \cite{unruh} (see also \cite{visser}). }.
The barotropic assumption gives
\begin{equation}
\rho_1=\frac{\partial \rho}{\partial p}p_1,
\end{equation}
After rearranging (\ref{ns}), we have
\begin{equation}
\label{lin}
\rho_1=-\frac{\partial \rho}{\partial p}\rho_0
\left(\partial_t \psi_1+\vec{v}_0\cdot \nabla\psi_1-\nu\nabla^2\psi_1\right),
\end{equation}
Substituting the above linearized equation of motion back into (\ref{co}), we obtain
\begin{eqnarray}
&&-\partial_t \bigg[\frac{\partial \rho}{\partial p}\rho_0
\left(\partial_t \psi_1+\vec{v}_0\cdot \nabla\psi_1-\nu\nabla^2\psi_1\right)\bigg]\nonumber\\
&&+\nabla\cdot\bigg[\rho_0
\nabla\psi_1-\frac{\partial \rho}{\partial p}\rho_0 \vec{v}_0
\left(\partial_t \psi_1+\vec{v}_0\cdot \nabla\psi_1-\nu\nabla^2\psi_1\right)\bigg]=0.
\end{eqnarray}
We can rewrite the above equation in a Klein-Gordon like form
\begin{equation}
\label{nkg}
\partial_{\mu}(\sqrt{-g}g^{\mu\nu}\partial_{\nu}\psi_1)=-\frac{\partial \rho}{\partial p}\rho_0 \nu
(\partial_t+\vec{v}_0\cdot\nabla)\nabla^2\psi_1,
\end{equation}
where the inverse metric is given by
\begin{equation}
\label{inverse}
 g^{\mu\nu}\equiv
\left(\frac{1}{\rho_0c_s}\right) \left[ \matrix{-1&\vdots&-v_0^j
\cr
               \cdots\cdots&\cdot&\cdots\cdots\cdots\cdots\cr
           -v_0^i &\vdots&-(c^2_s\delta^{ij}-v_0^{i}v_0^{j})  \cr }
\right], \label{nonrelativistic}
\end{equation}
and the local speed of sound is defined by
$c^{-2}_s=\frac{\partial \rho}{\partial p}$. The acoustic metric
$g_{\mu\nu}$ is given in (\ref{un}).

 What is different from before is that we have third order derivative term
of $\psi_1$ in the right hand of  (\ref{nkg}).
 The similar situation was discussed in \cite{visser} and Lorentz symmetry breaking was observed.
Clearly, by writing $\psi_1$ as $\rm \psi_1\sim exp(-i\omega
t+i\vec{k}\cdot \vec{x})$, one see that in the eikonal
approximation, the right hand of (\ref{nkg}) will be dominant, while
in the hydrodynamic limit, the right hand of (\ref{nkg})
can be ignored. In the following, we will work in long-wavelength
approximation and neglect those terms.

 In the cylindrical coordinate
$(r,\phi,z)$, the exact solution to Navier-Stokes equation yields
\cite{sullivan}
\begin{eqnarray}
\label{velocity}
v^{r}_{0}&=&-\alpha r+\frac{2\beta\nu}{r}\left(1-e^{-\frac{\alpha r^2}{2\nu}}\right),\\
v^{\phi}_{0}&=&\frac{\Gamma_0}{2\pi r}\frac{H(x)}{H(\infty)},\\
v^{z}_{0}&=&2 \alpha z\left[1-\beta e^{-\frac{ \alpha
r^2}{2\nu}}\right]
\end{eqnarray}
where
\begin{equation}
H(x)=H(\frac{ar^2}{2\nu})=\int^{x}_0 {\rm exp}\left(-s+\beta\int^s_0
\frac{1-e^{-\tau}}{\tau}d\tau\right)ds,
\end{equation}
$\Gamma_0$ is a constant that depends on the boundary condition,
$\nu$ denotes the shear viscosity, $\alpha>0$ and $\beta>1$ are free
parameters that can be fixed in experiments. By using the coordinate
transformation
\begin{eqnarray}
dt = d\tau +\frac{v^{r}_{0}}{{v^{r}_{0}}^2+{v^{z}_{0}}^2-c^2_s} dr
+\frac{v^{z}_{0}}{{v^{r}_{0}}^2+{v^{z}_{0}}^2-c^2_s} dz,\\
 d\phi = d\varphi + \frac{v^{r}_{0} v^{\phi}_{0}}{{v^{r}_{0}}^2+{v^{z}_{0}}^2-c^2_s}\frac{dr}{r}
+ \frac{v^{z}_{0}
v^{\phi}_{0}}{{v^{r}_{0}}^2+{v^{z}_{0}}^2-c^2_s}\frac{dz}{r}.
\end{eqnarray}
We obtain a non-spherically symmetric metric by rewriting (\ref{un})
as
\begin{eqnarray}
\label{nsm} &&ds^2=\frac{\rho_0}{
c_s}\bigg\{-(c^2_s-v^2_0)d\tau^2+\left(1-\frac{{v^{r}_{0}}^2}{{v^{r}_{0}}^2+{v^{z}_{0}}^2-c^2_s}\right)dr^2
+\left(1-\frac{{v^{z}_{0}}^2}{{v^{r}_{0}}^2+{v^{z}_{0}}^2-c^2_s}\right)dz^2\nonumber\\
&&-\frac{2v^{r}_{0}
v^{z}_{0}}{{v^{r}_{0}}^2+{v^{z}_{0}}^2-c^2_s}drdz-2
v^{\phi}_{0}rd\varphi d\tau+r^2 d\varphi^2\bigg\}.
\end{eqnarray}
It is worth noting that when $v_z=0$, we can recover the `` draining
bathtub'' metric obtained in \cite{visser}, that is to say
\begin{eqnarray}
&&ds^2=\frac{\rho_0}{
c_s}\bigg\{-(c^2_s-v^2_0)d\tau^2+\frac{c^2_s}{c^2_s-{v^{r}_{0}}^2}dr^2
+dz^2-2 v^{\phi}_{0}rd\varphi d\tau+r^2 d\phi^2\bigg\}.
\end{eqnarray}
The form of the metric obtained in (\ref{nsm}) is non-spherically
symmetric and even non axi-symmetric. There exists an event horizon
for the metric. The event horizon can be fixed by using the Null
surface equation\cite{hawking}
\begin{equation}
g^{\mu\nu}\frac{\partial f}{\partial x^{\mu}}\frac{\partial
f}{\partial x^{\nu}}=0,
\end{equation}
where $f=f(\tau,r,\varphi,z)=0$ denotes the hypersurface. Since the
radius of the event horizon would vary as a function of $\tau$ and
$z$, we find that the event horizon is determined by
\begin{equation}
\dot{r}_{H}=\left[{v^{r}_{0}}^2(r_{H})+{v^{z}_{0}}^2(r_{H})-c^2_s\right]
\bigg[{v^{r}_{0}}^2(r_{H})-r'_{H}{v^{r}_{0}}
(r_{H}){v^{z}_{0}}(r_{H})+r'^2_{H}{v^{z}_{0}}^2(r_{H})-c^2_s(1+r'^2_{H})\bigg]c^{-2}_s,
\end{equation}
where the dot and the prime denote the derivative with respect to
$\tau$ and $z$, respectively.

\section{Conclusion}
In summary, a new acoustic black hole metric has been derived from
the Abelian Higgs model. In the non-relativistic limit, the metric
can reduce to an acoustic black hole geometry given by Unruh. Since
the Abelian Higgs model describes high energy physics, our results
indicate that  acoustic black holes might be created in high energy
physical process, such as quark matters and neutron stars. The
dispersion relation was given and in the non-relativistic limit, we
can recover the linear dispersion relation. We have interpreted the
requirements of creating an acoustic black hole by using the
language of superconductor physics. Our results have  demonstrated
that although it would be very difficult to mimic the acoustic black
hole in type I superconductors,  people would be able to to discover
the acoustic black holes near the magnetic vortex in type II
superconductors . We developed an acoustic black hole with spiral
vortex geometry from the Navier-Stokes equation in classical
Newtonian fluids.

One point to be figured out is that we have used the vortices to
mimic the black hole.  Actually, when a superfluid flow exceeds the
Landau critical velocity, vortices can be created and excitations
with negative energy appear, which means that black hole
configurations are energetically unstable since the speed of sound
is also the the Landau critical velocity. However, this does not
mean that the supercritical flow of the superfluid vacuum is not
possible. In superfluid $\rm ^3He$ the Landau velocity for vortex
nucleation can be exceeded by several orders of magnitude, without
vortices creation. In \cite{vol}, the event horizon was proposed to
be realized in superfluid $\rm ^3He$ by arranging  the superfluid flow
velocity to be  constant but the ``sound velocity''  to  change  in space and
in some region becomes less than the flow velocity of the liquid. In
our above discussions, we have shown that in the ideal type II
superconductors, the ``sound velocity'' (\ref{sound})  indeed
depends on the superfluid density $\rho_0$, but the superconducting
(superfluid) flow velocity can be chosen  to be constant by varying
the current and the superfluid density to be $\rho_0$ simultaneously in
(\ref{v1}). Therefore, if the ``sound velocity'' changes in space,
superconducting (superfluid) flow velocity would exceed the ``sound
velocity'' in some regions.

 \vspace*{10mm} \noindent
 {\large{\bf Acknowledgments}}

\vspace{1mm}XHG thanks Hanyang university and CQUeST for warm
hospitality. XHG is indebted to Yunping Wang for helpful
discussions. This work was supported in part by the WCU project of
Korean Ministry of Education, Science and Technology
(R33-2008-000-10087- 0). The work of XHG was partly supported by
NSFC, China (No. 10947116), Shanghai Rising-Star Program(No.
10QA1402300) and Shanghai Leading Academic Discipline Project (No.
S30105). The work of SJS was supported in part by KOSEF Grant
R01-2007- 000-10214-0 and by the National Research Foundation of
Korea(NRF) grant funded by the Korea government(MEST)
(No.2005-0049409).

\renewcommand{\theequation}{A.\arabic{equation}}

\setcounter{equation}{0} \setcounter{footnote}{0}

\end{document}